%% file: main.tex
\documentclass[12pt,oneside,a4paper]{article}
\usepackage{authblk}
\usepackage{lineno}
\usepackage{lscape}
\usepackage{amsmath,amssymb,amsfonts}
\usepackage{algorithmic}
\usepackage{graphicx}
\usepackage{footnote}
\usepackage{url}
\usepackage{textcomp}
\usepackage{booktabs}
\usepackage{multirow}
\usepackage{bbding}
\usepackage[backgroundcolor=blue!10,bordercolor=gray]{todonotes}

\def\BibTeX{{\rm B\kern-.05em{\sc i\kern-.025em b}\kern-.08em
    T\kern-.1667em\lower.7ex\hbox{E}\kern-.125emX}}

\definecolor{lightgray}{rgb}{.9,.9,.9}
\definecolor{darkgray}{rgb}{.4,.4,.4}
\definecolor{purple}{rgb}{0.65, 0.12, 0.82}

\usepackage{color}
\usepackage{listings}
\usepackage{caption}
\usepackage{xcolor,colortbl}

\newcounter{nalg}[section] 
\renewcommand{\thenalg}{\arabic{nalg}} 
\DeclareCaptionLabelFormat{algocaption}{Listing \thenalg} 

\lstnewenvironment{algorithm}[1][] 
{   
    \refstepcounter{nalg} 
    \captionsetup{labelformat=algocaption,labelsep=colon} 
    \lstset{ 
        mathescape=true,
        frame=tB,
        numbers=left, 
        numberstyle=\tiny,
        basicstyle=\scriptsize, 
        keywordstyle=\color{black}\bfseries\em,
        keywords={,input, output, return, datatype, function, in, if, else, foreach, while, begin, end, } 
        numbers=left,
        xleftmargin=.04\textwidth,
        #1 
    }
}
{}

\makeatletter
    \let\@internalcite\cite
    \def\cite{\def\citeauthoryear##1##2{##1, ##2}\@internalcite}
    \def\shortcite{\def\citeauthoryear##1{##2}\@internalcite}
    \def\@biblabel#1{\def\citeauthoryear##1##2{##1, ##2}[#1]\hfill}
\makeatother

\begin{document}

\title{Employing Partial Least Squares Regression with Discriminant Analysis for Bug Prediction}

\author[1]{Rudolf Ferenc}
\author[1]{{Istv\'an Siket}}
\affil[1]{Department of Software Engineering, University of Szeged, Szeged, Hungary}
\author[1,2]{\\ {P\'eter Heged\H{u}s}}
\affil[2]{MTA-SZTE Research Group on Artificial Intelligence, University of Szeged, Szeged, Hungary 
}
\author[3]{{R\'obert Rajk\'o} \Envelope\,}
\affil[3]{Institute of Mathematics and Informatics, University of P\'ecs, P\'ecs, Hungary, rajko@gamma.ttk.pte.hu}

\maketitle

\newpage

\begin{abstract}
	Industry 4.0 (including Internet of Things, big data challenge, etc.) as industrial revolution system induces new contents in terms of software quality as well. 
	Forecasting defect proneness of source code has long been a major research concern.
	Having an estimation of those parts of a software system that most likely contain bugs may help focus testing efforts, reduce costs, and improve product quality.
	
	Many prediction models and approaches have been introduced during the past decades that try to forecast bugged code elements based on static source code metrics, change and history metrics, or both.
	However, there is still no universal best solution to this problem, as most suitable features and models vary from dataset to dataset and depend on the context in which we use them.
	Therefore, novel approaches and further studies on this topic are highly necessary.
	
	In this paper, we employ a chemometric approach -- Partial Least Squares with Discriminant Analysis (PLS-DA) -- for predicting bug prone Classes in Java programs using static source code metrics.
	PLS-DA is successfully applied within the field of chemometrics, but to our best knowledge, it has never been used before as a statistical approach in the software maintenance domain for predicting software errors.
	In addition, we have used rigorous statistical treatments including bootstrap resampling (we also used re-sampling and up-sampling for improving and balancing the training set, resp.) and randomization (permutation) test, and evaluation for representing the software engineering results.
	
	We show that our PLS-DA based prediction model achieves superior performances compared to the state-of-the-art approaches (i.e. F-measure of 0.44--0.47 at 90\% confidence level) when no data re-sampling applied and comparable to others when applying up-sampling on the largest open bug dataset, while training the model is significantly faster, thus finding optimal parameters is much easier.
	In terms of completeness, which measures the amount of bugs contained in the Java Classes predicted to be defective, PLS-DA outperforms every other algorithm: it found 69.3\% and 79.4\% of the total bugs with no re-sampling and up-sampling, respectively.
\end{abstract}
\newpage

\input{introduction}

\input{related}
\input{methodology}
\input{results}
\input{threats}
\input{conclusion}

\section*{Acknowledgment}
The presented work was carried out within the SETIT Project (2018-1.2.1-NKP-2018-00004)\footnote{Project no. 2018-1.2.1-NKP-2018-00004 has been implemented with the support provided from the National Research, Development and Innovation Fund of Hungary, financed under the 2018-1.2.1-NKP funding scheme.} and partially supported by grant TUDFO/47138-1/2019-ITM of the Ministry for Innovation and Technology, Hungary.
Support by the EU-supported Hungarian national grant GINOP-2.3.2-15-2016-00037 ``Internet of Living Things'' is also acknowledged.
Furthermore, P\'eter Heged\H{u}s was supported by the Bolyai J\'anos Scholarship of the Hungarian Academy of Sciences.

\input{bibliography}

\end{document}

%% file: introduction.tex
\section{Introduction}\label{sec:introduction}

For more than two decades, information systems mainly built up and programmed in Java~\cite{10.5555/861632}, however the correct operation of such systems highly depend on the quality of codes.  

Software bug prediction aims at forecasting defect-prone software components in order to help focus testing efforts and reduce post-release bugs.
Studies~\cite{4527256,6509481,challagulla2008empirical} agree that software product and process metrics are viable predictors of post-release defects.
However, we are far from an agreement on what the best predictors (static source code metrics, change/history metrics, etc.) and classification methods (e.g. regression, decision trees, support vector machines, neural networks, Bayesian method) are for the task.

In the literature, there is much on-going discussion regarding proper terminology and the purpose of classification. The use of such terms as classification or discrimination, depending on context and with a little additional explanation, indicate the construction of a specific type of classification rules. Basically, all discriminations cause classifications as well, so discrimination is a prerequisite for classification, however some classes/sets may remain undefined. That is why, one-class, two-class etc. classifications were introduced. In this paper we do not want to open this philosophical disputation, and we use two-class (or two-set) classification~\cite{ZHOU202016} based on a discriminating chemometric method.  

There is also a lot of debate around the usefulness of static source code metrics alone for defect prediction.
Many researchers found~\cite{fenton2000quantitative,shepperd1994critique} that static source code metrics capture too little information about source code to make them efficient defect predictors.
However, based on experiment results, Menzies and Greenwald~\cite{4027145} endorsed the use of static code attributes for predicting defects.
They argued that the pessimistic results of others are due to neglecting the following two important aspects of static source code based defect prediction:
\begin{itemize}
	\item The best attribute subset of defect predictors can change dramatically from dataset to dataset (i.e. there is no best predictor set).
	\item There is a large variance in the performances of the individual learning methods depending on the dataset and context they are used in (i.e. there is no best learning method).
\end{itemize}
The first point suggests that methods that are able to select the appropriate set of the partial defect predictors (i.e. static metrics) and combine them might be successful.
According to the second point, we need as many different methods for building defect prediction models as possible, since the more methods there are, the more likely it becomes that one will find an appropriate solution for a given context.
Catal and Diri~\cite{CATAL20091040} presented the first study which investigated the effects of dataset size, metrics set, and the feature selection techniques for software fault prediction problem. They concluded that the most crucial component in software fault prediction is the algorithm and not the metrics suite.

Adding into the equation the fact that static source code metrics are much easier to calculate (and only the current source code version is required) than change and code history metrics, we are convinced that static source code based defect prediction is still worthwhile to be investigated.

In this work, we propose a novel procedure for predicting defects in source codes using the static source code metrics based on an improved algorithm of partial least squares discriminant analysis (PLS-DA).
This method is widely used in chemometrics (a discipline that gathers chemical and related data evaluation techniques)~\cite{Geladi1986}, but gets much less attention in the software engineering domain.
We show that our PLS-DA based class level prediction model achieves superior performance compared to the state-of-the-art approaches (i.e. F-measure of 0.44--0.47 at 90\% confidence level) when no data re-sampling applied and comparable to others when up-sampling is applied on the largest open bug dataset we know~\cite{Ferenc:2018:PUB:3273934.3273936,FERENC2020110691,ferenc2020public}, while training the model is significantly faster, thus finding optimal parameters is much easier.
In terms of completeness, which measures the amount of bugs contained in the Java Classes predicted to be defective, PLS-DA outperforms every other algorithm: it found 69.3\% and 79.4\% of the total bugs with no re-sampling and up-sampling, respectively.
Additionally, the model is easily explainable (i.e. we can find the metrics that contribute the most to the prediction), as well as highly portable, as the model is basically a formula with a set of coefficients that can be used to make predictions anywhere using basic math operations (i.e. no need for deep learning or other run-time frameworks).

The main contributions of this work are:
\begin{itemize}
	\item A novel method for bug prediction based on PLS-DA, which is widely used in chemometrics, but not yet in the software maintenance domain.
	\item An empirical analysis on the largest available public bug dataset we know.
	\item To support open-science, we make the model implementation and metric normalization programs publicly available~\cite{Data4PLSDA}.
\end{itemize}

The rest of the paper is structured as follows.
In Section~\ref{sec:related}, we summarize the works that are related to ours.
We describe the data pre-processing steps we took and the construction of the PLS-DA based classification model in Section~\ref{sec:methodology}.
The evaluation and comparison of the proposed PLS-DA classification model is presented in Section~\ref{sec:results}.
We overview the possible threats to the validity of our work in Section~\ref{sec:threats}, and conclude the paper in Section~\ref{sec:conclusions}.

%% file: related.tex
\section{Related Work}\label{sec:related}

There is an abundance of works in the area of defect prediction.
However, to the best of our knowledge, we are the first to deploy a fully-fledged classification method based on the PLS-DA statistical method.

Based on the types of predictors they use, we can categorize the defect prediction models into two groups.
Models that use only static source code metrics as predictors (similarly to our approach), and models that might use so-called process metrics as well (e.g. code change/version control history metrics, like the number of changes of a particular code part during a time period).

Nagappan and Ball~\cite{Nagappan:2005:SAT:1062455.1062558} presented an empirical approach for the early prediction of pre-release defect density based on the defects found using static analysis tools.
The defects identified by two different static analysis tools were used to fit and predict the actual pre-release defect density for Windows Server 2003.
They showed that there exists a strong positive correlation between the static analysis defect density and the pre-release defect density determined by testing.
Thus, it is feasible to use static features to predict defects.
In contrast to their approach, we did not analyze the correlation between static issue detection tools and real bugs, rather we built a statistical model to predict defective Java Classes based on static source code metrics.

Based on their experiment results, Menzies and Greenwald~\cite{4027145} endorsed the use of static code attributes for predicting defects.
They investigated the prediction power of the classic static metrics (i.e. lines of code, Halstead metrics) for fault detection.
The authors found that there is a large variance in the performances of the individual learning methods depending on the dataset and context they are used in.
They achieved the best results with Naive Bayes classifier with log-transform.
They also showed that the best attribute subset of defect predictors can change dramatically from dataset to dataset.

Similarly to Menzies and Greenwald, Turhan and Bener~\cite{4385500} found the Bayesian method to be the best in static source code metric based defect prediction.
They applied it in conjunction with a multivariate approach and got promising results, recall around 80\% and precision around 30\%.

Kocagüneli et al.\cite{kocaguneli2009prest} introduced Prest, an open source tool for static analysis and bug prediction.
Compared to other open source prediction and analysis tools, Prest is unique in that it collects source code metrics and call graphs  in 5 different programming languages, and performs learning-based defect prediction and analysis.
Prest achieved an average of 32\% efficiency increase in testing effort.

Marchenko and Abrahamsson~\cite{marchenko2007predicting} studied the usefulness of two selected static analysis tools in defect rate prediction.
Five projects and 137 KLOC of the source code have been processed and compared to the actual defect rate.
As a result, a strong positive correlation with one of the tools was found.
It confirms the usefulness of a static code analysis tools as a way for estimating the amount of defects left in the product.

Encouraged by these results, we introduce a novel, statistical method (PLS-DA) for defect prediction with an extended set of static source code metrics as predictors.

Gray et al.~\cite{gray2009using} used a Support Vector Machine (SVM) classifier on eleven NASA data sets to predict faults based on static source code metrics.
After filtering and pre-processing the input data, they achieved 70\% accuracy in the classification.
We did not use SVM for classification, rather deployed a statistical method to create a novel bug classification approach.
However, we did compare our results to various other classical approaches, SVM being amongst them.
Moreover, we ran our experiment on the largest available open bug dataset containing more than 47 thousand entries.

In another work, Gray et al.~\cite{5596650} present an investigation that involves a manual analysis of the predictions made by Support Vector Machine classifiers using data from the NASA Metrics Data Program repository.
The findings show that the predictions are generally well motivated and that the classifiers were, on average, more `confident' in the predictions they made, which were correct.
A similar phenomenon was observed by Gyim\'othy et al.~\cite{GFS05}, who studied the bug prediction capabilities of individual static source code metrics.
They showed that static metrics (especially coupling, CBO) are good predictors for defects, but their performance is even better, when we count the total number of bugs found and not just the ratio of defective and non-defective Java Classes they identified.

Moreover, we not only analyze the binary classification performance of our new PLS-DA based method, but also the amount of total bugs found in the Java Classes identified as bugged.
Our results support that of Gyim\'othy et al., as our method performs better in identifying Java Classes with more than one bug.

Moser et al.~\cite{moser2008comparative}, Rahman et al.~\cite{rahman2014comparing}, and Nagappan et al.~\cite{4021972} present studies that compare the efficiency of bug prediction using static code metrics and history metrics.
They all conclude that history metrics tend to work better in some contexts, as well as being able to enhance the prediction power of static code metrics.

Okutan and Y{\i}ld{\i}z~\cite{okutan2014software} evaluated the effectiveness of various static and change metrics in defect proneness.
They concluded that response for (Java) class (RFC), lines of code (LOC), and lack of coding quality (LOCQ) are the most effective metrics, whereas coupling between objects (CBO), weighted method per class (WMC), and lack of cohesion of methods (LCOM) are less effective metrics on defect proneness.

Given the great number of contradicting research results, for example the usage of static metrics as predictors are encouraged~\cite{4027145} versus history metrics are better~\cite{moser2008comparative}; or CBO is a very good defect predictor~\cite{GFS05} versus CBO is less effective in defect prediction~\cite{okutan2014software}, further studies on this topic are still favorable.
Moreover, as static metrics are much easier to obtain for arbitrary systems, prediction models based only on static metrics might prove to be more flexible.

Recently, Li et al.~\cite{li2019evaluatingDBLP}, Pandey et al.~\cite{PANDEY2020113085} and Rhmann et al.~\cite{RHMANN2020419} have reported their results on using novel strategies and benchmarking studies for software fault prediction performance. Lately, Saifudin et al.~\cite{Saifudin_2019} as well as Goyal and Sardana~\cite{doi:10.1002/smr.2290} have strengthened the problem of class-imbalance and provided approaches to tackle that. 

There are a few works mentioning the Partial Least Squares (PLS) method in the context of defect prediction~\cite{ren2014software,luo2012asymmetric,tantithamthavorn2016automated,ramani2012predicting}.
However, in most of the cases, PLS is used as a substitute for PCA for dimension reduction and is not followed by an actual discriminant analysis and formal interpretation of the classification rules.
In contrast to these works, we deploy a fully fledged classification method by applying a discriminant analysis (DA) with PLS and show that the prediction results are comparable with other state-of-the-art techniques (even with deep learning for instance).

%% file: methodology.tex
\section{Approach}\label{sec:methodology}

In this work, we address the following research questions:
\begin{itemize}
	\item \textbf{RQ1}: Is PLS-DA feasible for predicting bugs in Java Classes based on static source code metrics?
	\item \textbf{RQ2}: What is the performance of PLS-DA in terms of completeness (number of bugs contained in the predicted bugged Java Classes)?
	\item \textbf{RQ3}: How does the performance of PLS-DA compare to other classification algorithms with and without data re-sampling?
\end{itemize}

Our approach to investigate these questions is detailed in the following sections.

\subsection{Dataset and predictors}

For creating, optimizing, and evaluating our statistical model, we used the Public Unified Bug Dataset for Java~\cite{Ferenc:2018:PUB:3273934.3273936,FERENC2020110691,ferenc2020public}.
It contains the data entries of 5 different public bug datasets (PROMISE~\cite{Sayyad-Shirabad+Menzies:2005}, Eclipse Bug Dataset~\cite{4273265}, Bug Prediction Dataset~\cite{5463279}, Bugcatchers Bug Dataset~\cite{Hall:2014:CSS:2668018.2629648}, and GitHub Bug Dataset~\cite{10.1007/978-3-319-42089-9_44}) in a unified manner.
Each row in the dataset marks a Java Class, for which 60 static source code metric values are provided.
The dataset contains the number of bugs in these Java Classes, which we used as the response variable (bugs).

We applied a manual process for feature reduction (i.e. removing redundant variables).
Some of the provided metrics were variants of the same measure, which we removed from further analysis.
For example, out of LOC (Lines of Code), LLOC (Logical Lines of Code), TLOC (Total Lines of Code), and TLLOC (Total Logical Lines of Code), we only kept LLOC.
We removed all the ``Total'' versions\footnote{Total means that the metric is calculated for the actual code element including all the contained elements, recursively.} of the metrics (TNM, TNOS, etc.) in general, as they clearly correlate with their counterparts.
Nonetheless, we did not strive for complete correlation analysis between metrics, as PLS, similarly to PCA, handles dependent features and reduces the dimension to find the appropriate combinations of the explanatory variables that describe most of the variance in the response variable.

The 35 static source code metrics provided as part of the dataset, which we kept after removing the above mentioned ones, are listed in Table~\ref{tab:metrics}.
The details of metric threshold (MAD Thresh. column) calculation is described in Section~\ref{sec:norm}.
We used these metrics as the explanatory variables in our model.

\input{data/metrics}

\begin{figure}[htb!]
\centering
\includegraphics[width=0.97\columnwidth]{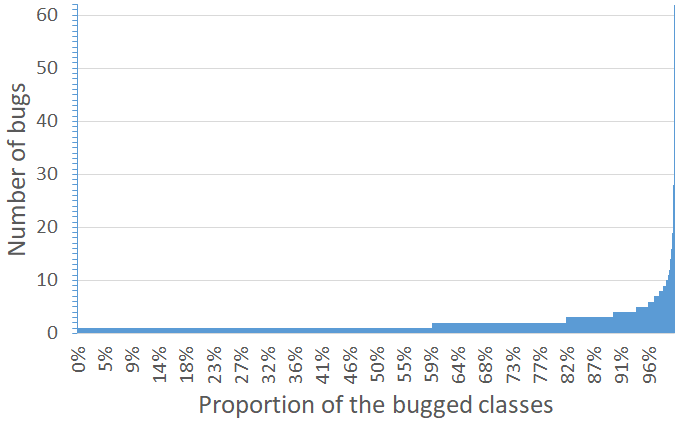}
\caption{The distribution of the bug numbers}
\label{fig:bug-dist}
\end{figure}

The dataset contains 47,618 Java Classes altogether, from which 8,780 contain at least one bug, while 38,838 are bug-free.
The total number of bugs recorded in the dataset is 17,365, which means that each bugged Java Class contains 1.98 bugs in average (with standard deviation of 2.39).
The distribution of the bug numbers within the bugged Java Classes is depicted in Figure~\ref{fig:bug-dist}.
As can be seen, almost 60\% of the bugged Java Classes contain only one bug, around 80\% of the Java Classes contain one or two bugs, etc.
Only about 6\% of the bugged Java Classes contain 5 or more bugs, but they contain a significant amount (i.e. 4,814, more than 27\%) of the total bugs.

\subsection{Data pre-processing}\label{sec:norm}

The appropriate pre-processing of the data before applying classification techniques is vital.
In our case, the metric values have entirely different ranges, for example, the clone coverage (CC) metric is a real value between 0 and 1, while, for example, the logical lines of code (LLOC) metric is a positive integer with no theoretical upper bound, but several orders of magnitude larger than CC.
Working with these values as is might result in poor performances regardless of the modeling method used.
Therefore, we normalized these values to fit uniformly into the [0, 1] range.

There are several approaches published in the literature for normalizing software metrics or deriving appropriate threshold values~\cite{Alves:2010:DMT:1912607.1913282,shatnawi2010finding,aniche2016satt,oliveira2014extracting,fontana2015automatic}.
The simplest method would be to take the maximum value in the bug dataset for each metric, and divide all the values with this maximum.
Even though this would transform the values into the desired range, it has some serious flaws.
In order to apply our prediction model and classify a new and unseen Java Class, we would need to normalize its metric values.
It might happen that the new Java Class has a larger metric value than that of the maximum in the bug dataset.
Therefore, we would run out of the [0,1] range unless we use this maximum value as a threshold and assign a normalized value of 1 for each metric value larger than the threshold.
Another problem is that the taken maximum might be an outlier value, which would cause the transformed values to be skewed (i.e. transformed values would be unnecessarily low).
Finally, the Java Classes in the unified bug dataset might not be representative, thus the above two issues might happen with large probabilities.

Inspired by the work of Alves et al.~\cite{Alves:2010:DMT:1912607.1913282}, we chose to implement our own benchmark-based threshold calculation that uses the median absolute deviation (MAD) measure of variability.
For an univariate data set $X_1$, $X_2$, \dots, $X_n$, the MAD is defined as the median of the absolute deviations from the data's median $\tilde{X} = median(X)$:
\begin{equation}\label{eq:1}
MAD = median(|X_i - \tilde{X}|).
\end{equation}

The MAD may be used similarly to how one would use the deviation for the average.
In order to use the MAD as a consistent estimator for the estimation of the standard deviation $\tilde{\sigma}$, one takes
\begin{equation}\label{eq:2}
\tilde{\sigma} = k \cdot MAD,
\end{equation}
where $k$ is a constant scale factor.

In contrast to Alves et al., we did not derive thresholds to represent a fixed percentage of the code, rather we let the variability of the metric data determine the cut point where we separate ``normal'' metric values from outliers.
In order to derive a threshold for each metric value (and thus achieving normalization by dividing the metric values with this threshold), we used a large benchmark of 217 mostly open-source Java systems (whose sizes range between 300 and 2 millions lines of code).
We derived the threshold values from the benchmark, which we then applied to normalize the metric values in the unified bug dataset.
The algorithm used for finding the appropriate threshold is shown in Listing~\ref{alg:mad3}.

The $MAD\_THRESHOLD$ function calculates the threshold value for a particular metric based on all the values of this metric in the benchmark (for all the Java Classes of the 217 systems).
First, we calculate the median (line 5) and the $\tilde{\sigma}$ value (line 6) of the metrics sample M (see Equation~\ref{eq:1} and \ref{eq:2}).
We chose 3 as the scale factor $k$.

\begin{algorithm}[caption={Threshold calculation using MAD}, label={alg:mad3}]
 function MAD_THRESHOLD
   input: list of metrics from the benchmark (M)
   output: the threshold value (T)
   begin
     med $\gets$ median(M)
     $\tilde{\sigma}$ $\gets$ 3 $\cdot$ MAD(M)
     if med = 0 $and$ $\tilde{\sigma}$ = 0
       M $\gets$ remove all but one zeros form M
       return MAD_THRESHOLD(M)
     end
     $T \gets$ largest $m_i \in M$
     if there are $m_i \in M$, $|m_i-med| \geq \tilde{\sigma}$
       $T \gets$ smallest $m_i \in M$, $|m_i-med| \geq \tilde{\sigma}$
     return T
   end       
\end{algorithm}

If the median and the $\tilde{\sigma}$ of the sample are both 0 (which indicates MAD is 0 as well), we removed all the 0 metric values from the sample keeping only one of them.
We then re-applied the threshold calculation with this modified metrics sample M (lines 7-10).
This is to handle the case when a lot of metric values are zeros, thus the distribution of non-zero values are distorted.
In other cases, the threshold value $T$ will be the smallest metric value from sample M that differs from the median of the samples with at least $\tilde{\sigma}$ in absolute value if there are such, or the largest metric value in M otherwise (lines 11-14).

Using the value $T$ calculated based on the algorithm above, we applied Equation~\ref{eq:3} to normalize the metric values in the bug dataset.
\begin{equation}\label{eq:3}
\overline{m_{bug}} = min(1, m_{bug}/T).
\end{equation}

This normalization means that we transform each metric value into the [0,1] interval,\footnote{Note, that all of the metric values are non-negative.} with a cut at threshold T, that is, all the metric values greater than $T$ will become a 1.
Such values are rare (i.e. around 8.6\% of the metric values in the entire bug dataset) and they are outliers anyway, however, they might carry important information regarding the classification.
Therefore, instead of removing, we truncated and included them as maximal values in our normalized data set.
As the threshold for normalization is calculated based on a very large and diverse universe of 217 software systems, we believe it is fairly representative.
We experimented with other types of normalizations as well, but by far this method resulted in the best classification performance.
The Python implementation of the MAD normalization is available in the online appendix.
Table~\ref{tab:metrics} lists the resulted threshold values for each metric.

\begin{figure*}
	\centering
	\includegraphics[width=\textwidth,height=0.45\textheight,trim = 15mm 65mm 0mm 90mm, clip]{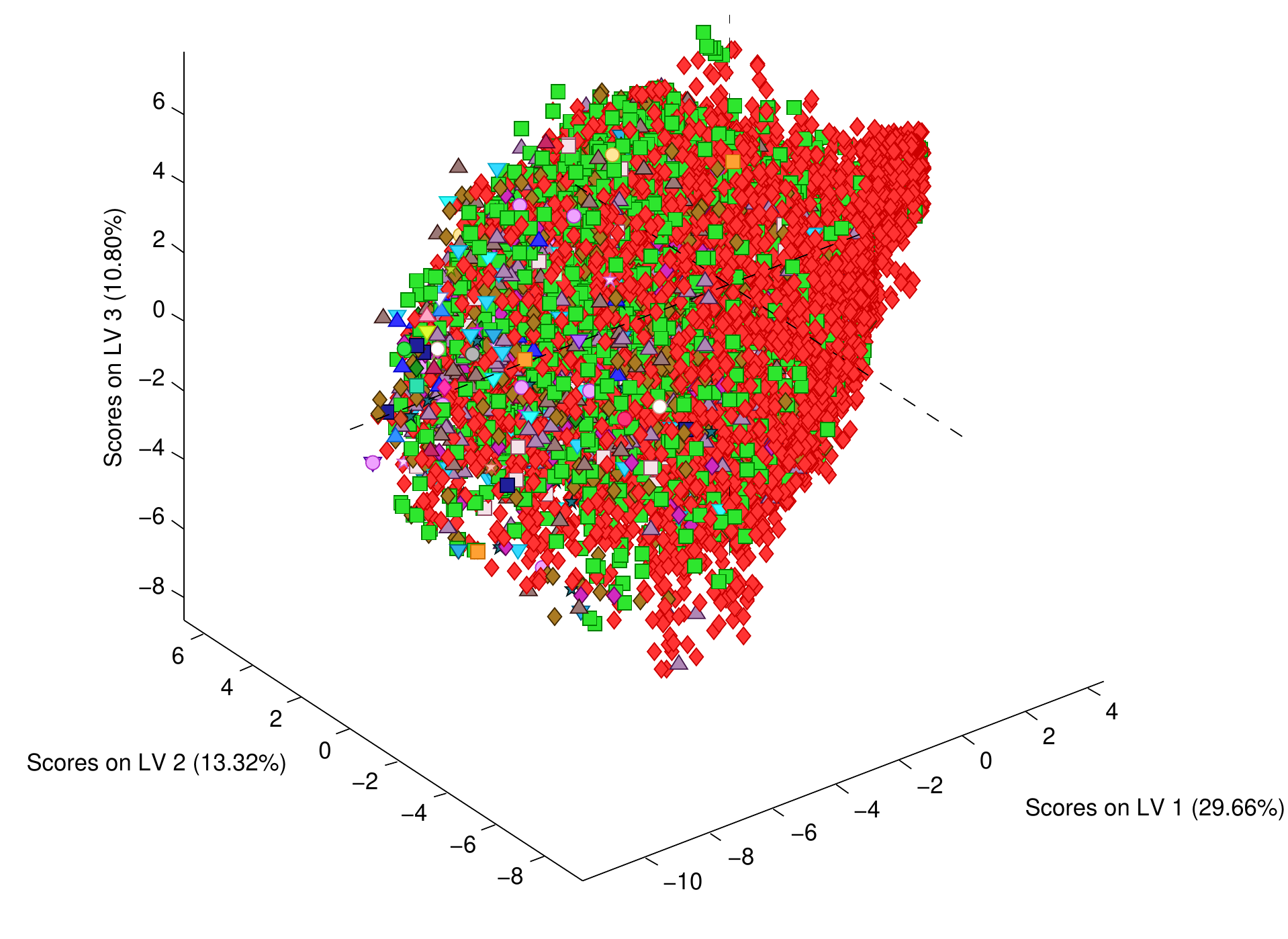}
	\caption{Preliminary result obtained by using PLS\_Toolbox PLSDA function for the overall dataset - the score plot (3D abstract space) of non-bugged Java Classes (red diamonds), and bugged Java Classes (green squares for containing only 1 bug, and other markers for depicting more than one bug)}
	\label{fig:PLSDA_Plstlb_LV7_3D}
\end{figure*} 

\subsection{Model building and evaluation}~\label{sec:meth-plsda}

Partial least squares (PLS) regression is a multiple inverse least squares method, which is basically biased as it treats multi-collinea\-rity in that way~\cite{martens1992multivariate}.
PLS is mainly applied in chemometrics (a discipline that gathers chemical and related data evaluation techniques)~\cite{Geladi1986}, therefore, we use the terms accustomed to that field at least in its introduction.

A univariate calibration model can only provide accurate results if the measured signal does not have contributions from other sources~\cite{Bro_multivariatecalibration:2003}. 
Hence, only the analyte (important chemical component) of interest must contribute to the measured signal. 
If other analytes (interferents) contribute to the signal, the results will be distorted. 
What is worse is that there is no way to detect that we obtained incorrect results by observing the univariate signal alone.
Therefore, the adequateness of the result becomes a matter of belief (as well as experience and external validation). 
Multivariate models are more adequate than univariate models: they can handle situations that cannot be handled univariately, in particular, it is possible to incorporate interferents and to have automatic outlier detection when building or using a model~\cite{Bro_multivariatecalibration:2003}.

The general underlying model of multivariate PLS is 
\begin{equation}\label{eq:plsXscore}
\mathbf{X} = \mathbf{T} \, \mathbf{P}^{T} + \mathbf{E},
\end{equation}
\begin{equation}\label{eq:plsYscore}
\mathbf{Y} = \mathbf{U} \, \mathbf{Q}^{T} + \mathbf{F},
\end{equation}
\begin{equation}\label{eq:plsOverall}
\mathbf{Y} = \mathbf{X} \, \mathbf{B} + f(\mathbf{E},\mathbf{F}),
\end{equation}
where $\mathbf{X}$ is an $\displaystyle n\times m$ matrix of predictors ($n$ and $m$ are the number of cases and variates/features/metrics, respectively), $\mathbf{Y}$ is an $\displaystyle n\times p$ matrix of responses ($p$ is the number of variables/classes/sets); $\mathbf{T}$ and $\mathbf{U}$ are $\displaystyle n\times l$ matrices that are, respectively, projections of $\mathbf{X}$ (the \textit{X-score}, component or factor matrix) and projections of $\mathbf{Y}$ (the \textit{Y-score}); $\mathbf{P}$ and $\mathbf{Q}$ are, respectively, $\displaystyle m\times l$ and $\displaystyle p\times l$ orthogonal loading matrices ($l$ is the number of latent variables corresponding to the PLS abstract space where the projections are); and matrices $\mathbf{E}$ and $\mathbf{F}$ are the error terms, assumed to be independent and identically distributed random normal variables. 
The decompositions of $\mathbf{X}$ and $\mathbf{Y}$ are made so as to maximize the covariance between $\mathbf{T}$ and $\mathbf{U}$ generating the $\displaystyle m\times p$ coefficient matrix $\mathbf{B}$, and $f(\mathbf{E},\mathbf{F})$ is the error term which is a function of matrices $\mathbf{E}$ and $\mathbf{F}$. 

In addition to regression, PLS can be used for discrimination or classification as well. 
Partial least squares discriminant analysis (PLS-DA) is a multivariate inverse least squares discrimination method used to classify samples~\cite{Brereton:2014}. 
The y-block in a PLS-DA model indicates which samples are in the class(es)/set(s) of interest through either:
(a) a column vector of class/set numbers indicating class/set assignments,
(b) a matrix of one or more columns containing a logical zero (= not in class/set) or one (= in class/set) for each sample (row).
Thus, in fact, PLS-DA is a PLS with an extra step, i.e., classification by using a threshold for predicted y-values. It means that a PLS-DA model as well as PLS-DA results inherit all methods and properties from the conventional PLS objects.

At first, we calculated PLS-DA models and predictions using a user-friendly Matlab program package, PLS\_Toolbox~\cite{wise-plstoolbox-2007,pls_toolbox:2016} because of its menu-driven controlling facility.
We could investigate the preliminary results on the overall unified bug dataset, to see the positions of Java Classes as abstract points, see Figure~\ref{fig:PLSDA_Plstlb_LV7_3D}.
According to Equation~\ref{eq:plsXscore}, LVs are latent variables: the rows of matrix $\mathbf{T}$ will be the coordinates, thus we can get the projections of $\mathbf{X}$ onto the space spanned by the columns of matrix $\mathbf{P}$. 
This figure shows that the classification task is really hard, because the hulls of the abstract points belonging to different classes/sets cannot be easily identified and discriminated. 

Unfortunately, the PLS-DA implementation in PLS\_Toolbox was too slow due to the tremendous amount of administrative calculations it performs.
Therefore, we have developed and used a much faster PLS-DA script independently from PLS\_Toolbox. 
According to the literature, there is no obvious way to choose the fastest and most accurate algorithm~\cite{Andersson_9_PLS1_algs:2009,Indahl_geom_props_PLS1:2014,Bjorck_fast_and_stable_PLS:2017}. 
Thus, we had to find the right balance between speed and accuracy, and chose the {\it{bidiag2stab}} method~\cite{Indahl_geom_props_PLS1:2014} for our implementation.
For tuning the model parameters and finding the best possible classification, we performed many model training runs, thus a very fast PLS core implementation was essential.

With our PLS-DA Matlab script, we generated a classification using data-splitting of 80\% training, 10\% validation and 10\% test sets.
The PLS-DA model was built using the training sets, after which the latent variables were selected using the validation sets, and the model performance was evaluated on the test sets. 
We performed 1000 random repetitions (i.e. bootstrap resampling) of model building for evaluation and 10-fold cross-validation when compared to other algorithms. 
For the test sets we kept the original ratio of the bugged and the non-bugged Java Classes in the original data set.

For evaluating the performance of the algorithm, we used the measures defined by equations~\ref{eq:prec}-\ref{eq:mcc}, i.e., precision, recall and F1-score for both positive and negative views (calculated for bugged and non-bugged Java Classes), and MCC (Matthews Correlation Coefficient).

\begin{equation}\label{eq:prec}
Precision_p = \frac{\text{TP}}{\text{TP + FP} }
\end{equation}
\begin{equation}\label{eq:rec}
Recall_p = \frac{\text{TP}}{\text{TP + FN} }
\end{equation}
\begin{equation}\label{eq:fmes}
F1p = \frac{2 \cdot Precision_p\cdot Recall_p}{Precision_p + Recall_p}
\end{equation}
\begin{equation}\label{eq:prec_n}
Precision_n = \frac{\text{TN}}{\text{TN + FN} }
\end{equation}
\begin{equation}\label{eq:rec_n}
Recall_n = \frac{\text{TN}}{\text{TN + FP} }
\end{equation}
\begin{equation}\label{eq:fmes_n}
F1n = \frac{2 \cdot Precision_n\cdot Recall_n}{Precision_n + Recall_n}
\end{equation}
\begin{equation}\label{eq:mcc}
MCC = \frac{\text{TP}\times\text{TN}\,-\,\text{FP}\times\text{FN}}{\sqrt{\text{(TP\,+\,FP)(TP\,+\,FN)(TN\,+\,FP)(TN\,+\,FN)}}}
\end{equation}

%% file: data/metrics.tex
\begin{table}[htbp]
  \centering
  \caption{Static source code metrics used for classification of Java Classes}
    \begin{tabular}{l|l|l|r}
    \textbf{Category} & \textbf{Metric} & \textbf{Description} & {MAD Thresh.}\\
    \hline
    \hline
    Clone & CC    & Clone Coverage & 1.0\\
    metrics& CCL   & Clone Classes & 7\\
    & CCO   & Clone Complexity & 23\\
    & CI    & Clone Instances & 7\\
    & CLC   & Clone Line Coverage & 1.0\\
		& CLLC  & Clone Logical Line Coverage & 1.0\\
		& LDC   & Lines of Duplicated Code & 81\\
		& LLDC  & Logical Lines of Duplicated Code & 74\\
    \hline
		Cohesion & LCOM5 & Lack of Cohesion in Methods 5 & 5\\
		metrics & & & \\
		\hline
		Complexity & NL    & Nesting Level & 6\\
		metrics & NLE   & Nesting Level Else-If & 6\\
		& WMC   & Weighted Methods per Class & 23\\
		\hline
		Coupling & CBO   & Coupling Between Object classes & 12\\
		metrics & CBOI  & CBO Inverse & 6\\
		& NII   & Number of Incoming Invocations & 12\\
		& NOI   & Number of Outgoing Invocations & 11\\
		& RFC   & Response set For Class & 30\\
		\hline
		Documentation & AD    & API Documentation & 0.909\\
		metrics & CD    & Comment Density & 0.612\\
		& CLOC  & Comment Lines of Code & 22\\
		& PDA   & Public Documented API & 6\\
		& PUA   & Public Undocumented API & 7\\
		\hline
		Inheritance & DIT   & Depth of Inheritance Tree & 6\\
		metrics & NOA   & Number of Ancestors & 6\\
		& NOC   & Number of Children & 7\\
		& NOD   & Number of Descendants & 7\\
		& NOP   & Number of Parents & 4\\
		\hline
		Size & LLOC  & Logical Lines of Code & 111\\
		metrics & NA    & Number of Attributes & 17\\
		& NG    & Number of Getters & 6\\
		& NM    & Number of Methods & 34\\
		& NOS   & Number of Statements & 51\\
		& NPA   & Number of Public Attributes & 12\\
		& NPM   & Number of Public Methods & 29\\
		& NS    & Number of Setters & 12\\
		\hline
    \end{tabular}%
  \label{tab:metrics}%
\end{table}%

%% file: results.tex
\section{Results}\label{sec:results}

\subsection{Performance of the model for bugged/non-bugged Java Class discrimination}

According to Figure~\ref{fig:PLSDA_Plstlb_LV7_3D}, we can be in doubt that the classification procedure could be successful ever, at least if we use linear discriminating function. 
That is why we have checked the randomness of the PLS-DA result.
A so-called randomization (permutation) test is suitable here, it has been successfully applied to solve related problems in chemometrics~\cite{Faber_Rajko_Perm_test:2006,Wiklund_etal_Rand_test:2007,Faber_Rajko_Perm_test:2007}. 
The randomization test in regression consists of building models repeatedly after randomly permuting the rows in $\mathbf{Y}$.
The random permutation destroys the relationship that might be there.
Only correlations by chance between $\mathbf{X}$ and $\mathbf{Y}$ remain.
The Matlab function 'comodite.m' (COnditional MOdel DImensionality TEst for univariate PLS)~\cite{Faber_Rajko_Perm_test:2006,Wiklund_etal_Rand_test:2007,Faber_Rajko_Perm_test:2007} implemented by us yielded to catastrophically bad results: none of the latent variables was significant, confirming the hopeless situation depicted in Figure~\ref{fig:PLSDA_Plstlb_LV7_3D}.
However, the CoMoDiTe tests the regression capacity of PLS,
 but here we want to use PLS for discrimination (i.e., using PLS-DA).
In these cases, the variable $\mathbf{y}$ (when $\mathbf{Y}$ contains only one column it can be treated as a vector) is a dummy variable, containing labels for classification instead of real numbers, that is why the variance of this dummy variable is synthetic.

\begin{figure}[htb!]
	\centering
	\includegraphics[width=\columnwidth]{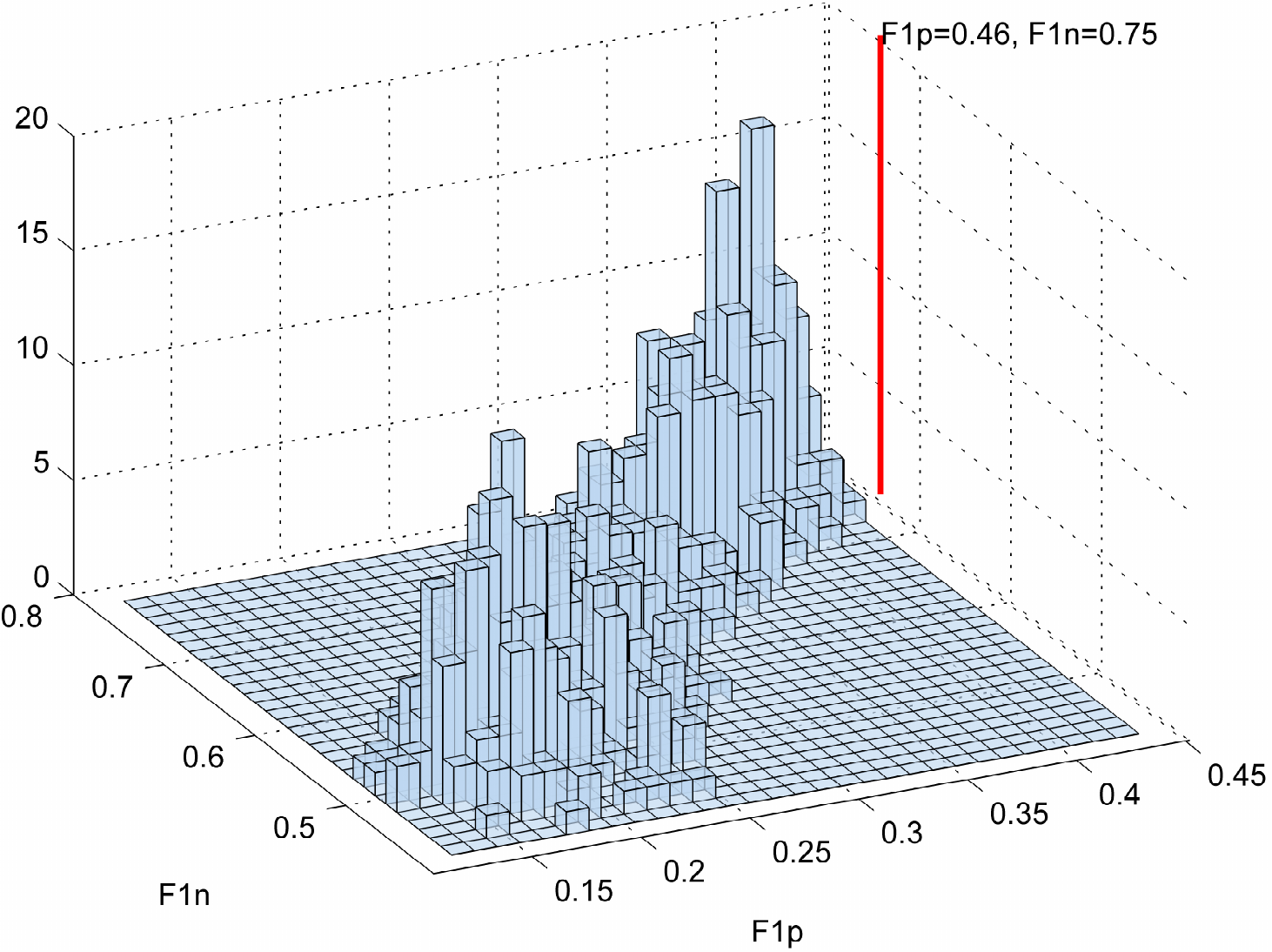}
	\caption{Permutation test for verifying that the PLS-DA result cannot be randomly generated}
	\label{fig:PLSDA_permtest}
\end{figure}

Therefore, we performed a permutation test for PLS-DA as a classification method.
The results are shown in Figure~\ref{fig:PLSDA_permtest}, where we can see that the PLS-DA result (F1p=0.46 and F1n=0.75) based on the original dataset is significantly separated from the bulk of the randomly generated permutation data.
In this case, the significance-level can definitely be considered as 0; if the PLS-DA result (F1p=0.46 and F1n=0.75) fell in the area covered by the histogram, we would have been able to get larger significance-level, but again, if it was less than, say, 0.05, the conclusion would be the same.
For the permutation test, a very fast implementation of the PLS-DA was essential of course. 

\begin{figure}[htb!]
	\centering
	\includegraphics[width=0.9\columnwidth]{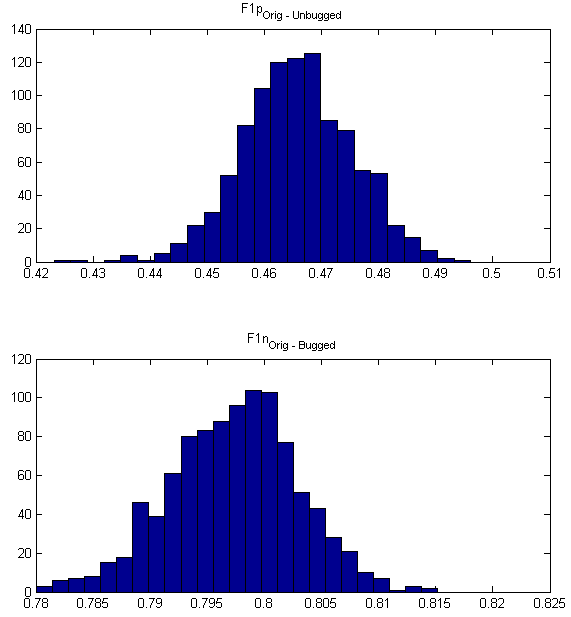}
	\caption{The distributions of F1p and F1n over 1000 random repetitions of model building}
	\label{fig:F1p_and_F1n}
\end{figure}

Table~\ref{tab:conf-matrix} illustrates a typical result of a 1000 times repetition using 80-10-10 percents splits summarized by the confusion matrix.
The overall performance measures are Precision$_p$ of 0.35 (SD 0.008), Recall$_p$ of 0.695 (SD 0.019), F1p of 0.466 (SD 0.010), Precision$_n$ of 0.911 (SD 0.005), Recall$_n$ of 0.709 (SD 0.009), and F1n of 0.797 (SD 0.006).
Figure~\ref{fig:F1p_and_F1n} depicts the distribution of the F1p and F1n values over the 1000 repetitions.
We can see that the performance measures are very stable (the standard deviation of the values are low), meaning that the results of PLS-DA do not depend much on the random selection of the training instances.
Another observation is that the original bugged/non-bugged ratio in the test sample weakened the performance of PLS-DA: the presence of outnumbered non-bugged Java Classes resulted in too many false positive decisions.

\medskip
\noindent
\fbox{%
  \parbox{0.975\columnwidth}{%
   \textbf{Answer to RQ1:} The proposed PLS-DA based bug prediction model gives meaningful results, thus its application in the domain of software defect prediction is feasible.
  }%
}

\input{data/confusion-matrix.tex}

\subsection{Performance of the model for finding the total number of bugs}

\begin{figure}[htb!]
	\centering
	\includegraphics[width=0.9\columnwidth]{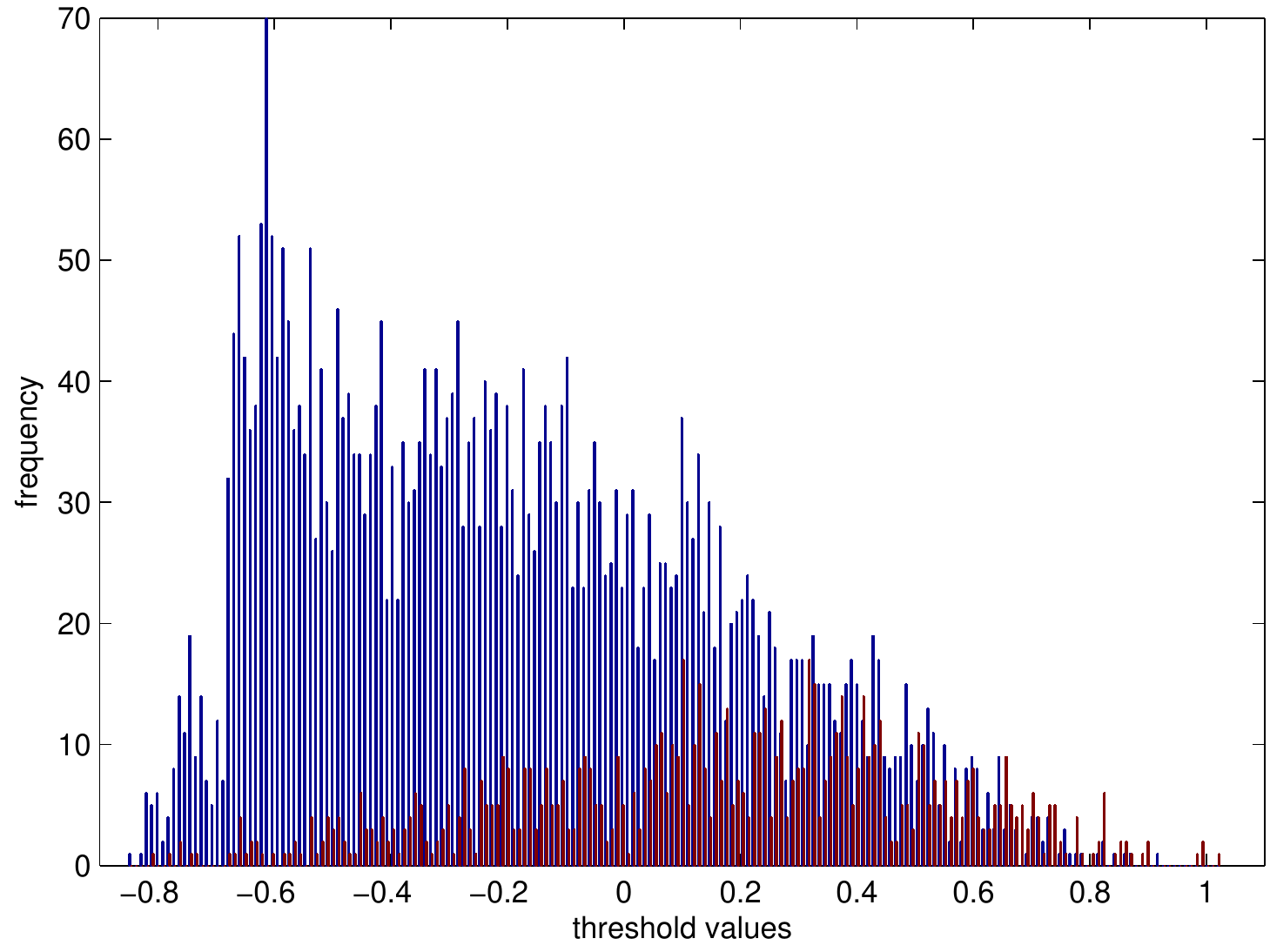}
	\caption{Threshold overlapping: the blue columns represents the non-bugged, while the claret columns the bugged subset}
	\label{fig:Threshold_overlapping}
\end{figure}

Figure~\ref{fig:Threshold_overlapping} shows a typical interim distributions of the PLS-DA calculation.
The default threshold value was 0 (originally the label was 1 for the bugged Java Classes ignoring the more than one bug cases, and -1 for non-bugged subset), thus if PLS-DA predicted negative values, the non-bugged decision was made as classification, and for larger than 0 predictions, the bugged decision was chosen. 
The predicted values are distributed with serious overlapping: the blue columns should have been at the negative part, and the claret columns should have been at the positive part.

\begin{figure}[htb!]
	\centering
	\includegraphics[width=0.9\columnwidth]{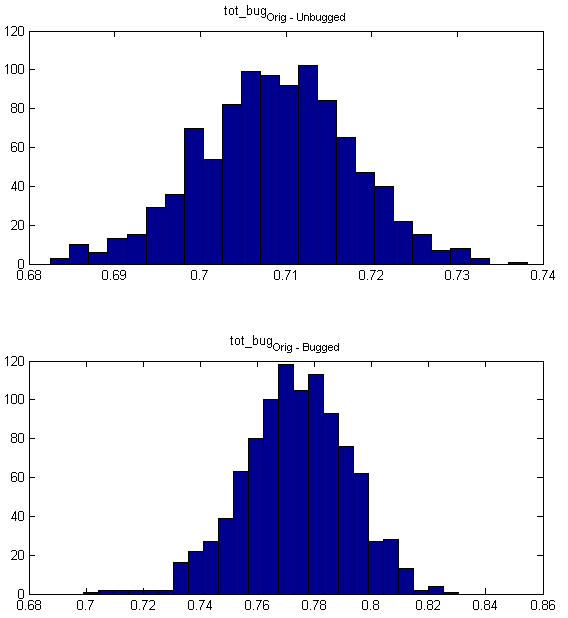}
	\caption{The distribution of the total bug ratio (i.e. completeness) over 1000 random repetitions}
	\label{fig:total_bug_ratio}
\end{figure}

Nonetheless, in terms of completeness, which measures the amount of bugs contained in the Java Classes predicted to be defective, PLS-DA achieved a value of 69.3\% (standard deviation 0.037) that is significantly higher than the recall of 0.593.
This means that despite our method finds only less than 60\% of the bugged Java Classes, these Java Classes contain almost 70\% of the total number of bugs.
Therefore, lower number of bugs occurred at the false positive part, and the larger number of bugs could be identified in the true positive part.
Figure~\ref{fig:total_bug_ratio} shows the distribution of completeness measures over 1000 random model buildings.

\medskip
\noindent
\fbox{%
  \parbox{0.975\columnwidth}{%
   \textbf{Answer to RQ2:} PLS-DA performs well in terms of completeness; it tends to find bugged Java Classes with large number of bugs correctly, and fails more when the bug numbers are close to 1.
  }%
}

\input{data/model-comparison}

\subsection{Comparison of the results with other classification methods}
To put the performance of our proposed PLS-DA based bug prediction model into context, we compared its results with 8 other well-known and widely used machine learning algorithms.
We used the TensorFlow~\cite{abadi2016tensorflow} framework\footnote{\url{https://www.tensorflow.org/}} for DNN and scikit-learn\footnote{\url{https://scikit-learn.org/stable/}} for the other 7 algorithms:
\begin{itemize}
	\item Deep neural network (DNN)
	\item Random forest algorithm (Forest)
	\item K-nearest neigbors algorithm (KNN)
	\item Support Vector Machine (SVM)
	\item Decision tree (Dec.~tree)
	\item Logistic regression (Log.~reg.)
	\item Linear regression (Lin.~reg.)
	\item Bayesian Network (Bayes)
\end{itemize}
All the algorithms have been run with their best hyper-parameters, which we found with an extensive grid-search algorithm on a separated 10\% of the bug dataset (dev set).
We compared the results both on the original imbalanced dataset and universally applying a 50\% up-sampling for each model training.

Table~\ref{tab:model-comp} displays the performance measures (the classical precision~(eq.~\ref{eq:prec}), recall~(eq.~\ref{eq:rec}), F1p~(eq.~\ref{eq:fmes}), their symmetric counterparts defined for non-bugged Java Classes (eq.~\ref{eq:prec_n}), (eq.~\ref{eq:rec_n}), (eq.~\ref{eq:fmes_n}), Matthews correlation coefficient~(eq.~\ref{eq:mcc}), and completeness) of this eight algorithms together with our proposed PLS-DA method without re-sampling and a 50\% up-sampling, respectively (best values are marked in bold and highlighted). 
All these measures except for completeness can be calculated with the help of the confusion matrix, from the true positive, false positive, true negative, and false negative values of the predictions that are also displayed in the table.
Below each algorithm, the standard deviations of the measures are shown (we applied 10-fold cross-validation for each algorithm).

The first remarkable observation is that up-sampling has a very significant effect on the algorithm performances.
When no re-sampling has been used, our proposed new PLS-DA based bug prediction method dominated in many performance measures.
It achieved the best recall (and symmetrically, precision for non-bugged Java Classes), F1p score and completeness.
PLS-DA had the highest number of true positive and false negative hits as well.
In terms of precision (and symmetrically, recall for non-bugged Java Classes), true negative and false positive hits SVM was at the top, while DNN had the highest MCC.
What stands out is the completeness measure of PLS-DA.
Our proposed algorithm found almost 70\% of all the bugs while achieving the best F1p score trained on the original, highly imbalanced dataset.
Only the Bayesian method was even close to that (completeness of 0.685), all the other algorithms performed much worse on the imbalanced dataset.

Upon applying a 50\% up-sampling, the performance measures changed significantly.
The general tendency was that the precision values decreased while the recall values increased for all algorithms.
For PLS-DA and Bayesian method, this change was marginal, the F1p scores remained almost the same.
The rest of the algorithms improved dramatically in terms of F1p, DNN achieving the best value of 0.525.
Our PLS-DA algorithm kept the best recall (and non-bugged precision), completeness, true positive and false negative counts.
In all other respect, random forest algorithm performed the best with this re-sampling strategy.
The PLS-DA's strength of identifying the bugged Java Classes with most bugs became more evident with re-sampling, PLS-DA found almost 80\% of the total bugs, more than 7\% more than the second performer DNN.

Regarding Precision$_n$, Recall$_n$, and F1n, which measure the same as their $_p$ counterparts with the difference that they are defined from the perspective of non-bugged Java Classes instead of the bugged ones, we can observe much higher values.
It means that the models perform better in identifying non-bugged Java Classes.
From this respect, the algorithms are much closer to each other, PLS-DA performs the best in terms of Precision$_n$ (0.915), while random forest has the best Recall$_n$ (0.865) and F1n (0.881) values.

We can conclude that PLS-DA will find the most of the bugged Java Classes compared to other algorithms (particularly those having the largest number of bugs) but for that, it pays the price of the highest false positive (FP) rate.
But it can do so even for highly imbalanced training datasets without any re-sampling, which is a clear advantage over all the other methods except for the Bayesian algorithm.
The fact that PLS-DA finds the most bugs implicates that there are bug occurrences that are predicted only by our proposed technique.

In addition, the PLS-DA based classification method is highly tunable by various parameters (see Section~\ref{sec:meth-plsda}).
By adjusting the value of the classification threshold parameter or optimizing the training for another measure (not F1p that we used), we can practically achieve an arbitrary distribution of precision and recall values maintaining approximately the same F1p measure.
This is a very big advantage of the proposed method, and since model training is very fast compared to, for example, DNNs or random forest, one can easily explore the best suitable value for a given bug prediction context.
Moreover, the results are explainable (we can find the metrics that contribute most to the prediction) as well as highly portable, as the model is basically a formula with a set of coefficients that can be used to make predictions anywhere using basic math operations (i.e. no need for deep learning or other run-time frameworks).

The Std.Dev. rows show the standard deviations of the values above them.
We ran all the algorithms with 10-fold cross-validation and calculated the average values based on the confusion matrices.
In general, all the deviations are very low, which means that the algorithms provide a very stable result, meaning that the partition of the data into training, validation and test set has minor effect on the final model performances.

\medskip
\noindent
\fbox{%
  \parbox{0.975\columnwidth}{%
   \textbf{Answer to RQ3:} The performance of PLS-DA is comparable to that of the other classification algorithms. It is one of the best methods when no re-sampling is applied on the training data, but has the highest recall and completeness even if we use an 50\% up-sampling.
  }%
}

%% file: data/confusion-matrix.tex
\begin{table}[htbp]
  \centering
	\setlength\tabcolsep{3pt}
  \caption{Confusion matrix}
    \begin{tabular}{cr|ll}
          &       & \multicolumn{2}{c}{Predicted ($\pm$Std.Dev.)} \\
          &       & Bugged & Not Bugged \\
    \hline
    \multirow{2}[2]{*}{Determined} & Bugged & TP=609.8 ($\pm$16.9) & FN=268.2 ($\pm$16.9) \\
          & Not Bugged & FP=1131.2 ($\pm$34.5) & TN=2752.8 ($\pm$34.5) \\
    \hline
    \end{tabular}%
  \label{tab:conf-matrix}%
\end{table}%

%% file: data/model-comparison.tex
\begin{landscape}\centering
\vspace*{\fill}
	\begin{table}[htb!]
		\centering
		\scriptsize
		\setlength\tabcolsep{0.6pt}
		\caption{Performance measures of the various learning methods without re-sampling and using 50\% up-sampling}
			\begin{tabular}{l|rrrrrrrrrrrrrrrrrrrrrrrr}
			Alg.  & \multicolumn{2}{c|}{Precision$_p$} & \multicolumn{2}{c|}{Recall$_p$} & \multicolumn{2}{c|}{F1p} & \multicolumn{2}{c|}{Precision$_n$} & \multicolumn{2}{c|}{Recall$_n$} & \multicolumn{2}{c|}{F1n} & \multicolumn{2}{c|}{MCC} & \multicolumn{2}{c|}{Compl.} & \multicolumn{2}{c|}{TP} & \multicolumn{2}{c|}{TN}& \multicolumn{2}{c|}{FP} & \multicolumn{2}{c}{FN}\\
			 & \multicolumn{1}{c|}{NO} & \multicolumn{1}{c|}{UP} & \multicolumn{1}{c|}{NO} & \multicolumn{1}{c|}{UP} & \multicolumn{1}{c|}{NO} & \multicolumn{1}{c|}{UP} & \multicolumn{1}{c|}{NO} & \multicolumn{1}{c|}{UP} & \multicolumn{1}{c|}{NO} & \multicolumn{1}{c|}{UP} & \multicolumn{1}{c|}{NO} & \multicolumn{1}{c|}{UP} & \multicolumn{1}{c|}{NO} & \multicolumn{1}{c|}{UP} & \multicolumn{1}{c|}{NO} & \multicolumn{1}{c|}{UP} & \multicolumn{1}{c|}{NO} & \multicolumn{1}{c|}{UP} & \multicolumn{1}{c|}{NO} & \multicolumn{1}{c|}{UP} & \multicolumn{1}{c|}{NO} & \multicolumn{1}{c|}{UP} & \multicolumn{1}{c|}{NO} & \multicolumn{1}{c}{UP}\\
			\hline
			\hline
			PLS-DA & 0.374 & 0.338 & \cellcolor{gray!50}\textbf{0.593} & \cellcolor{gray!50}\textbf{0.721} & \cellcolor{gray!50}\textbf{0.458} & 0.460 & \cellcolor{gray!50}\textbf{0.894} & \cellcolor{gray!50}\textbf{0.915} & 0.774 & 0.681 & 0.830 & 0.780 & 0.300 & 0.300 & \cellcolor{gray!50}\textbf{0.693} & \cellcolor{gray!50}\textbf{0.794} & \cellcolor{gray!50}\textbf{520.9} & \cellcolor{gray!50}\textbf{633.1} & 3006.6 & 2642.7 & 876.4 & 1240.3& \cellcolor{gray!50}\textbf{357.1} & \cellcolor{gray!50}\textbf{244.9}\\
			Std.Dev. & 0.013 & 0.010 & 0.037 & 0.028 & 0.011 & 0.009 & 0.006 & 0.006 & 0.022 & 0.025 & 0.011 & 0.015 & 0.013 & 0.013 & 0.037 & 0.026 & 32.7 & 24.4 & 86.3 & 96.3 & 86.3 & 96.3 & 32.7 & 24.4\\
			\hline
			\hline
			DNN & 0.592 & 0.443 & 0.337 & 0.643 & 0.429 & \cellcolor{gray!50}\textbf{0.525} & 0.863 & 0.910 & 0.947 & 0.817 & 0.903 & 0.861 & \cellcolor{gray!50}\textbf{0.360} & 0.403 & 0.424 & 0.723 & 295.7 & 564.3 & 3679.9 & 3175.2 & 203.9 & 708.6 & 582.3 & 313.7\\
			Std.Dev. & 0.021 & 0.020 & 0.038 & 0.022 & 0.030 & 0.014 & 0.004 & 0.004 & 0.005 & 0.016 & 0.004 & 0.008 & 0.021 & 0.018 & 0.035 & 0.021 & 33.7 & 19.3 & 35.3 & 62.6 & 35.3 & 62.6 & 33.7 & 19.3\\
			\hline
			Forest & 0.676 & \cellcolor{gray!50}\textbf{0.485} & 0.228 & 0.562 & 0.341 & 0.521 & 0.848 & 0.897 & 0.975 & \cellcolor{gray!50}\textbf{0.865} & \cellcolor{gray!50}\textbf{0.907} & \cellcolor{gray!50}\textbf{0.881} & 0.326 & \cellcolor{gray!50}\textbf{0.404} & 0.317 & 0.657 & 200.1 & 493.4 & 3787.9 & \cellcolor{gray!50}\textbf{3360.5} & 95.9 & \cellcolor{gray!50}\textbf{523.3} & 677.9 & 384.6\\
			Std.Dev. & 0.022 & 0.008 & 0.015 & 0.021 & 0.018 & 0.013 & 0.001 & 0.004 & 0.001 & 0.003 & 0.001 & 0.002 & 0.017 & 0.015 & 0.020 & 0.018 & 13.1 & 18.3 & 8.8 & 13.2 & 8.8 & 13.2 & 13.1 & 18.3\\
			\hline
			KNN & 0.564 & 0.438 & 0.331 & 0.589 & 0.418 & 0.502 & 0.862 & 0.899 & 0.942 & 0.829 & 0.900 & 0.862 & 0.341 & 0.375 & 0.406 & 0.663 & 290.9 & 516.8 & 3659.2 & 3219.8 & 224.6 & 664.0 & 587.1 & 361.2\\
			Std.Dev. & 0.019 & 0.011 & 0.014 & 0.021 & 0.014 & 0.014 & 0.001 & 0.005 & 0.002 & 0.004 & 0.002 & 0.003 & 0.016 & 0.018 & 0.020 & 0.023 & 12.5 & 18.1 & 16.2 & 16.4 & 16.2 & 16.4 & 12.5 & 18.1\\
			\hline
			SVM & \cellcolor{gray!50}\textbf{0.715} & 0.440 & 0.074 & 0.537 & 0.134 & 0.483 & 0.826 & 0.890 & \cellcolor{gray!50}\textbf{0.993} & 0.845 & 0.902 & 0.867 & 0.191 & 0.355 & 0.145 & 0.636 & 64.9 & 471.4 & \cellcolor{gray!50}\textbf{3857.9} & 3283.2 & \cellcolor{gray!50}\textbf{25.9} & 600.6 & 813.1 & 406.6\\
			Std.Dev. & 0.029 & 0.012 & 0.006 & 0.024 & 0.009 & 0.014 & 0.001 & 0.005 & 0.000 & 0.008 & 0.001 & 0.004 & 0.010 & 0.017 & 0.023 & 0.023 & 4.9 & 21.4 & 3.8 & 32.6 & 3.8 & 32.6 & 4.9 & 21.4\\
			\hline
			Dec. tree & 0.558 & 0.425 & 0.305 & 0.570 & 0.394 & 0.487 & 0.857 & 0.895 & 0.945 & 0.825 & 0.899 & 0.859 & 0.323 & 0.355 & 0.383 & 0.646 & 267.8 & 500.9 & 3671.8 & 3205.2 & 212.0 & 678.6 & 610.2 & 377.1\\
			Std.Dev. & 0.021 & 0.011 & 0.025 & 0.027 & 0.025 & 0.013 & 0.003 & 0.005 & 0.002 & 0.010 & 0.002 & 0.005 & 0.024 & 0.016 & 0.032 & 0.035 & 22.0 & 24.0 & 14.5 & 40.7 & 14.5 & 40.7 & 22.0 & 24.0\\
			\hline
			Log. reg. & 0.586 & 0.420 & 0.203 & 0.551 & 0.301 & 0.477 & 0.843 & 0.891 & 0.968 & 0.828 & 0.901 & 0.858 & 0.270 & 0.343 & 0.305 & 0.652 & 178.2 & 483.8 & 3757.7 & 3216.8 & 126.1 & 667.0 & 699.8 & 394.2\\
			Std.Dev. & 0.028 & 0.011 & 0.017 & 0.023 & 0.022 & 0.015 & 0.002 & 0.005 & 0.001 & 0.005 & 0.001 & 0.004 & 0.023 & 0.019 & 0.026 & 0.023 & 15.2 & 20.3 & 9.2 & 20.9 & 9.2 & 20.9 & 15.2 & 20.3\\
			\hline
			Lin. reg. & 0.628 & 0.415 & 0.144 & 0.555 & 0.234 & 0.475 & 0.835 & 0.891 & 0.981 & 0.823 & 0.902 & 0.856 & 0.240 & 0.340 & 0.226 & 0.655 & 126.4 & 487.6 & 3808.9 & 3196.8 & 74.9 & 687.0 & 751.6 & 390.4\\
			Std.Dev. & 0.042 & 0.011 & 0.014 & 0.022 & 0.020 & 0.014 & 0.001 & 0.005 & 0.001 & 0.006 & 0.001 & 0.004 & 0.025 & 0.018 & 0.023 & 0.024 & 12.0 & 19.7 & 8.9 & 21.6 & 8.9 & 21.6 & 12.0 & 19.7\\
			\hline
			Bayes & 0.354 & 0.346 & 0.577 & 0.605 & 0.439 & 0.440 & 0.889 & 0.892 & 0.762 & 0.742 & 0.821 & 0.810 & 0.287 & 0.288 & 0.685 & 0.708 & 506.9 & 531.2 & 2960.2 & 2880.3 & 923.6 & 1003.5 & 371.1 & 346.8\\
			Std.Dev. & 0.008 & 0.008 & 0.017 & 0.016 & 0.011 & 0.010 & 0.002 & 0.004 & 0.003 & 0.006 & 0.003 & 0.004 & 0.014 & 0.014 & 0.018 & 0.016 & 15.3 & 14.5 & 22.8 & 24.6 & 22.8 & 24.6 & 15.3 & 14.5\\
			\hline
			\end{tabular}%
		\label{tab:model-comp}%
	\end{table}%
\vfill
\end{landscape}

%% file: threats.tex
\section{Threats to Validity}\label{sec:threats}

There are several threats to the validity of our work that we took measures to protect against.
The quality of the unified bug dataset we used for training the model highly impacts our conclusions.
We could not validate the entries it contains, but the dataset has been assembled from several other, publicly available datasets, which are used in many papers for testing bug prediction methods.
Therefore, the threat posed by errors in the underlying bug dataset is negligible.

We trained the models based on the static source code metrics provided as part of the unified bug dataset.
Since we did not calculate these measure by ourselves, we had to trust the quality of this measure published as part of the dataset.
To mitigate this threat, we randomly picked and cross-checked some of the values, which we found to be precise.

We applied our own Matlab implementation of the PLS-DA method due to performance reasons, which may contain some programming issues.
To reduce the likelihood of programming errors, we thoroughly tested our Matlab script.
We also ran some comparison tests with the official PLS\_Toolbox to make sure that both implementations give the same results.

%% file: conclusion.tex
\section{Conclusions and Future Work}\label{sec:conclusions}

In this paper, we proposed a PLS-DA based statistical method for classifying bugged or non-bugged Java Classes based on static source code metrics.
The Partial Least Squares method is widely used in chemometrics, but to the best of our knowledge, we are the first to employ a fully-fledged classification method for bug prediction based on PLS-DA.

Besides describing the technical details of our new method, we also performed an empirical study using the largest open bug dataset we know (i.e. containing more than 47 thousand Java Classes and 17 thousand bugs).
We achieved an F-measure (F1p) of 0.466 with recall of 0.695 and precision of 0.35.
We also compared these results to 8 other state-of-the-art learning approaches.
Without applying any re-sampling, PLS-DA based bug prediction method dominated in many performance measures.
It achieved the best recall (and symmetrically, precision for non-bugged Java Classes), F1p score and completeness (number of bugs contained in the found bugged Java Classes).
When a 50\% up-sampling has been applied, PLS-DA produced comparable results to those of other algorithms, keeping the best recall and completeness measures.
Our proposed algorithm found almost 70\% and 80\% of all the bugs when no re-sampling and 50\% up-sampling was applied, respectively.

The linear and some non-linear model building based classification methods suffer from the collinearity problem of the dependent variables, some of them can treat it partially using the embedded algorithms, but PLS-DA can naturally handle collinearity implicitly because of the used subspace shrinking (dimensional reduction of the PLS abstract space) concept.

Overall, our results are comparable to those of the existing approaches with clear strengths and more flexible parameter adjustments.
It is very easy to experiment with parameter settings, as the training of the model is fast.
Moreover, the results of the model are explainable (i.e. we can point out those metric values that contribute the most to the prediction) and the model is highly portable, as it is basically a formula with a set of coefficients that can be used to make predictions anywhere using basic math operations (i.e. no need for deep learning or other run-time frameworks). 
To support open-science, we made the model implementation and normalization programs publicly available~\cite{Data4PLSDA}.

There are numerous further research directions in connection with the proposed PLS-DA method that we plan to explore.
During the validation phase, a target measure can be set up to optimize, which we chose to be F1p in this version.
In addition, one can even adjust the threshold values for making the decision of whether a Java Class is bugged or not, which we also fixed in this work.
In the future, we plan to upgrade our PLS-DA based method to allow the easy tuning of these parameters and perform further empirical studies on their effects.

We learned that the non-linearity of the classification model highly improves its performance (due to the complex separation of the bugged and non-bugged Java Classes).
Although PLS handles non-linearity to some extent, it needs further development for improving its performance.
The speedy implementation offers a convenient environment for trying several ideas because the runs can be done within a short time frame.

%% file: bibliography.tex
\bibliographystyle{acm}
\bibliography{bibl}